# Thomas-Fermi model yields condensed phase of molecular metallic oxygen

Yuri Kornyushin

*Maître Jean Brunschvig Research Unit, Chalet Shalva, Randogne, CH-3975*

Calculations, performed in Thomas-Fermi approximation, show that the energy of a condensed phase of molecular metallic oxygen is lower by 496 kJ/mol than that of an insulator oxygen phase. The insulator phase is separated from a metallic one by energetic barrier of 1165 kJ/mol. This barrier could be overcome by means of electromagnetic irradiation with quantum energy of 12.071 eV or a bit higher. During such a transition the amount of energy equal to 496 kJ/mol is released. It should be extracted from a sample to secure the existence of a metallic phase.

## 1. Introduction

Experimental evidence of existence of molecular fluid metallic oxygen was reported in [1]. Here calculations in Thomas-Fermi approximation are performed to analyze the situation.

Regarding ionization problem it is assumed tacitly that the charge of the delocalized electrons is distributed uniformly throughout a sample [2]. The ionization energy is usually regarded to be equal to that of an isolated atom/molecule. It is well known that a uniform distribution of delocalized electrons in a sample is not equilibrium in the presence of the ions [2]. The charge redistributes itself, causing screening of the field of the ions [2]. This relaxation of the charge distribution leads to the decrease in the energy of a system. As a result the effective ionization energy decreases also.

When $N$ randomly distributed ions, charged with elementary charge $e$, and uniformly distributed charge of delocalized electrons are present, the electrostatic energy of a system consists of the energy of the ions [3,4], $e^2N/2\varepsilon r_0$ (here $\varepsilon$ is dielectric constant and $r_0$ is some small distance, referring to the ion size), the energy of their charge distributed uniformly throughout a sample [4], the energy of the uniformly distributed charge of the delocalized electrons, and the energy of the interaction of the two uniform charges. The energy of the uniformly distributed ions charge, the energy of the uniformly distributed charge of the delocalized electrons, and the energy of their interaction, all three of them, annihilate together because the uniformly distributed negative and positive charges compensate each other. What's left is the energy of the bare ions, $e^2N/2\varepsilon r_0$.

When the delocalized electrons relax to equilibrium, screening the ions, the electrostatic energy of a sample decreases. This means that the effective ionization energy of the atoms/molecules decreases also. Screening cuts off the long-range field, that is, leads to a decrease in the energy of electrostatic field.

All this is taken into account here to analyze the possibility of existence of metallic molecular condensed oxygen phase.

## 2. Model

Let us consider an ion, charged with elementary positive charge $e$. Usually microscopic electrostatic field around charged ions is neglected when the ionization energy is concerned. The ionization energy is considered to be equal to that of a separate atom/molecule [2]. Recently it was shown that the microscopic electrostatic field contributes very essentially to thermodynamic properties [3]. The matter is that when a separate atom/molecule is ionized, the electron becomes a plane wave in an infinite or very large space. This plane wave does not posses electrostatic field and electrostatic energy. Before the ionization of the atom/molecule we have electrostatic fields of the charges of the ion and the localized electron. These fields contribute to the formation of the ground state energy of the atom/molecule. The field of the electron charge does not act on electron itself, only Coulombic electrostatic field of the ion acts. But the energy of the electrostatic field of the charge of a localized electron contributes to formation of the energy of a ground state. After the ionization of a separate atom/molecule we have a Coulombic electrostatic field around the ion. Its energy contributes to the formation of a quantum ground state and its energy also as was mentioned. When we have many randomly distributed atoms/molecules ionized, we have many delocalized electrons in a sample. The charge of these electrons redistributes itself to equilibrium distribution, causing screening of Coulombic field of the point charge of the ion and decrease in the electrostatic energy of a system.

We consider here the ions as point charges, and the delocalized electrons like a negatively charged gas.

The electrostatic field around a separate positive ion submerged, into the gas of delocalized electrons, is as follows [2]:

$$\varphi = (e/\varepsilon r)\exp{-gr}, \qquad (1)$$

where $r$ is the distance from the center of the ion and $1/g$ is the screening radius.

The electrostatic energy of this field is smaller than that of a bare ion. So the electrostatic energy of a system is also smaller as a result of the screening. The electrostatic energy of a separate ion with electrostatic field [Eq. (1)] is the integral over the volume of a sample of its gradient in square, multiplied by $\varepsilon/8\pi$. The lower limit of the integral on $r$ should be taken as $r_0$, a very small value as was mentioned above. Otherwise the integral diverges. Taking into account that the value of the volume of a sample is usually very large (comparative to the ion volume) and performing calculation, we come to the following expression for the electrostatic energy of a separate ion [3]:

$$W = 0.5e^2[(r_0^{-1} + 0.5g)/\varepsilon]\exp{-2gr_0}. \qquad (2)$$

As $gr_0$ is very small comparative to unity, Eq. (2) yields the following expression for the electrostatic energy of a separate ion [3]:

$$W = (e^2/2\varepsilon r_0) - 0.75(e^2g/\varepsilon). \qquad (3)$$

Here $e^2/2\varepsilon r_0$ is the electrostatic energy of the bare ion in a dielectric medium. When $g$ is small this term is the only one in the right-hand part of Eq. (3).

The decrease in the electrostatic energy of a separate ion due to the screening is $-0.75e^2g/\varepsilon$. Electrostatic energy of $N$ randomly distributed ions is just $NW$ [4]. It should be mentioned here also that the screened fields of the separate ions [Eq. (1)] do not overlap in the object considered. Calculation shows that these fields in a metal with one delocalized electron per atom/molecule start overlapping when the current carrier concentration $n$ is larger than $4.1\times10^{26}$ cm$^{-3}$. So in any case the electrostatic energy of $N$ ions is just $NW$ as was mentioned before.

To ionize the first atom/molecule the initial ionization energy $W_0$ is needed. When delocalized electrons are present, the energy of a system decreases by $0.75e^2g/\varepsilon$, due to the relaxation, accompanied by the screening of the ion. The effective ionization energy of the ion is correspondingly as follows:

$$W_e = W_0 - 0.75(e^2g/\varepsilon). \qquad (4)$$

This problem was considered earlier in [5]. In [5] the ionization energy of impurities in semiconductors was discussed. The effective ionization energy was calculated in [5] for the case when the radius of a ground state is much smaller than the screening radius. The result was as follows:

$$W_e = W_0 - (e^2 g/\varepsilon). \tag{5}$$

This corresponds to the first term in the right-hand part of Eq. (2).

While calculating the effective ionization energy here no assumption, concerning the relation between the sizes of a ground state and the screening radius, was made.

### 3. Dielectric constant and screening

Let us consider now a degenerate gas of delocalized electrons with concentration $n$. We assume here that the concentration of delocalized electrons is equal to that of atoms/molecules. For this case the screening radius $1/g$ is Thomas-Fermi radius, defined by the following relation:

$$g^2 = 6\pi e^2 n/\varepsilon E_F, \tag{6}$$

where $E_F = (\hbar^2/2m)(3\pi^2 n)^{2/3}$ is Fermi energy.

To calculate the effective ionization energy in this case, the kinetic energy of the delocalized electrons should be taken into account. The kinetic energy of the degenerate delocalized electrons per unit volume of a sample is as follows [2]:

$$T = 0.6 n E_F(n). \tag{7}$$

The effective ionization energy per one molecule in the case regarded is accordingly as follows:

$$W_e = W_0 + (T/n) - 0.75(e^2 g/\varepsilon). \tag{8}$$

The decrease in the energy of a system and in the effective ionization energy is inversely proportional to $\varepsilon^{3/2}$. For $\varepsilon = 16$, $\varepsilon^{3/2} = 64$. The factor $1/\varepsilon^{3/2}$ can be very essential.

When $\varepsilon$ is large enough there are two options of a screening: the screening radius $1/g$ is essentially smaller than the average distance between the ions, $(6/\pi n)^{1/3}$, and vice versa. The first option corresponds to an essentially smaller energy of a system. So this option should be realized. When the screening radius $1/g$ is essentially smaller than the average distance between the ions, $(6/\pi n)^{1/3}$, one should consider $\varepsilon = 1$, because there is no polarization of the medium where the non-zero field is found. This corresponds to a much larger decrease in the energy of a system due to the screening. For $n = 8 \times 10^{21}$ cm$^{-3}$ we have $(6/\pi n)^{1/3} = 6.2 \times 10^{-8}$ cm and $1/g = 8.2 \times 10^{-9}$ cm for $\varepsilon = 1$. For $n = 10^{24}$ cm$^{-3}$ we have $(6/\pi n)^{1/3} = 1.24 \times 10^{-8}$ cm and $1/g = 3.67 \times 10^{-9}$ cm for $\varepsilon = 1$. We see that in the considered concentration range $1/g$ is essentially smaller than average distance between the ions. So we should accept $\varepsilon = 1$ while calculating energy decrease.

## 4. Condensed metallic oxygen

When the effective ionization energy is zero or negative, full ionization of the atom/molecule occurs.

Let us perform further calculations for molecular oxygen. In this case $W_e$ = 12.071 eV [6]. Let us introduce another variable, $x$, according to the following relation:

$$n = 10^{24} x^6. \tag{9}$$

Then the effective ionization energy $W_e$ is zero or negative when

$$x^4 \leq 1.3418x - 0.5504. \tag{10}$$

This happens when $0.43745 \leq x \leq 0.90076$, which corresponds to

$$7.0076 \times 10^{21} \text{ cm}^{-3} \leq n \leq 5.3414 \times 10^{23} \text{ cm}^{-3}. \tag{11}$$

In the terms of the density, $d$, Eq. (11) yields $0.3727 \text{ g/cm}^3 \leq d \leq 28.37 \text{ g/cm}^3$.

It should be mentioned here that when the total ionization occurs the delocalized electron concentration $n$ is equal to the number of the molecules per unit volume.

The upper limit in Eq. (11) is essentially larger than observed atomic density in regular objects of condensed matter. This means that in case the metallic oxygen phase exists, the range, defined by Eq. (11), is the one is relevant. In this range the screening radius is essentially smaller that the average distance between the ions.

The insulator phase and the metallic one are separated by energetic barrier, 12.071 eV = 1165 kJ/mol. This barrier could be overcome by means of electromagnetic irradiation of the surface of the liquid oxygen. In case such a transition occurs, the amount of the energy equal to 496 kJ/mol is released, according to Eq. (8). This energy should be extracted from a sample to secure the existence of a metallic phase.

## 3. Discussion

For elements and molecules with high enough initial ionization energy $W_0$ the insulator-metal transition discussed here does not exist. The inequality, equivalent to Eq. (10), has no solution in this case accordingly. Molecular oxygen was chosen here because the ionization energy of the oxygen molecule is not too high. So there exists a rather wide range of the densities for which the insulator-metal transition may take place, as follows from Thomas-Fermi approximation.

Transition discussed here does not lead to the change in the number of molecules and their distribution. So from this part there is no change in the entropy of a system as a result of transition. Delocalized electrons are degenerate. The range of Fermi temperatures is from 15520 K to 281523 K according to Eq (11) and definition of Fermi energy, given immediately after Eq. (6). That is we have a very strong degeneracy here. Every electron sits on its level and the contribution of the delocalized electrons to the change in the entropy of a system is negligible.

Possible influence of the dielectric constant of the substance was neglected also in the calculations performed. It should be noted, however, that the average distance between the molecules $(6/\pi n)^{1/3}$ is larger from 3 to 6 times than the screening radius $1/g$ in the interval considered [see Eqs. (6) and (11)]. So it looks like every molecule should be treated as a separate one without introducing dielectric constant into the calculations.

Calculations performed here were first published in [7] for molecular oxygen gas.